\documentclass[a4paper]{panl}
\usepackage{cite}
\usepackage{wrapfig}
\usepackage{graphicx}
\usepackage{amssymb}
\usepackage{amsfonts}
\usepackage{amsmath}
\usepackage{longtable}
\usepackage{rotating}
\usepackage{lscape}
\usepackage{epsfig}
\usepackage{multirow}
\usepackage{slashed}

\originalTeX
\begin{document}

\title{
Production of single isolated photons in the \\
Parton Reggeization Approach }
\maketitle
\authors{
A.\,Chernyshev$^{a,}$\footnote{E-mail: aachernyshoff@gmail.com},
V.\,Saleev$^{a,b,}$\footnote{E-mail: saleev.vladimir@gmail.com}}
\setcounter{footnote}{0}

\from{$^{a}$\,Samara University, Samara, Russia}

\from{$^{b}$\,Joint Institute for Nuclear Research, Dubna}

\begin{abstract}

In the article, we study the processes of single isolated photon
production at the LHC energies in the framework of the Parton
Reggeization Approach taking into account LO ${\cal
O}(\alpha^1\alpha_S^0)$ and NLO${}^\star$ ${\cal
O}(\alpha^1\alpha_S^1)$ contributions, the last one includes only
tree--level corrections. Reggeized amplitudes are constructed
according to the effective field theory formalism for multi--Regge
kinematics processes suggested by L.N.~Lipatov. To avoid the double
counting between tree--level corrections and unintegrated parton
distribution functions, a subtraction scheme is introduced. The
results of calculations are compared with experimental data.

\end{abstract}
\vspace*{6pt}

\noindent
PACS: 12.38.Bx, 13.85.Qk, 14.70.Bh

\label{sec:intro}
\section{Introduction}

Single photons produced in the hard parton scattering (so called {\em direct photons})
are considered as an important probe of the perturbative QCD,
i.e. fixed order Collinear Parton Model (CPM) calculations,
since these are single--scale processes.
Additionally, this process is a clean chanel
to study the gluon parton distribution functions (PDFs) through the dominant
leading order (LO) QCD <<Compton--like>> scattering at the LHC energies and
in the future SPD NICA experiment~\cite{SPD}. It is difficult to
experimentally extract the contribution of direct photons, so the production
of {\em prompt photons} (direct plus fragmentation photons) or
{\em isolated} ones (prompt photons plus using isolation condition
discussed in the Sec.~\ref{sec:num}) is usually studied.

The processes of single direct photon production also provide an
opportunity to study non--collinear parton dynamics using the High Energy
Factorization (HEF) approach which deal with the Regge limit of QCD where the
following hierarchy of the light--cone components is
valid\footnote{We use Sudakov decomposition of $4$--momenta:
$\forall p \colon p = p_L + p_T$, $p_L = \left( p^+ n_- + p^- n_+ \right) / 2$,
with light--cone basis vectors $n^\pm \colon (n^\pm, n^\mp) = 2$,
$\pm$--componenets are obtained as projections $p^\pm = (p, n^\pm)$. We
also use notation $\slashed{p} = p_\mu \gamma^\mu$.}:
$q^\mp \ll |{\bf q}_T| \sim \mu \sim q^\pm \ll \sqrt{s}$,
here $\mu \sim E_T^\gamma$ is a hard scale.
In the study, we use Parton Reggeization Approach (PRA)~\cite{PRA1,NS:2020},
which is a gauge--invariant version of the HEF. Such processes for the first time
were studied in the PRA in Refs.~\cite{Saleev:2008,Saleev:2009} within the LO PRA.
It is interesting to study they once again within NLO${}^\star$ approximation of the
PRA using the new unintegrated parton distribution functions (uPDFs) proposed in
Ref.~\cite{NS:2020}.

The paper has the folllowing structure: in the Sec.~\ref{sec:pra} we shortly
discuss the main ingridients of NLO${}^\star$ calculations in the PRA and parton
subprocesses in the direct photon production.
The results of calculation are presented in the
Sec.~\ref{sec:num}, as well as the comparison with experimental data.
Our conclusions are summarized in the Sec.~\ref{sec:con}.

\section{Parton Reggeization Approach}
\label{sec:pra}

In the PRA, hadronic cross section is given as a convolution of hard scattering
coefficient for Reggeized initial states with uPDFs~\cite{PRA1,NS:2020},
which is valid in the leading and next--to--leading logarithmic
approximation~\cite{kT1}:
\begin{equation}
d \sigma =
\Phi(x_1, t_1, \mu^2) \otimes {\cal H}(x_i, t_i, \mu^2) \otimes \Phi(x_2, t_2, \mu^2)
+ {\cal O}\left( \frac{\Lambda^\#}{\mu^\#}, \frac{\mu^2}{S} \right), \label{eq1}
\end{equation}
where $t_i = {\bf q}_{T_i}^2$.
Off--shell initial partons in the PRA are interpreted as
Reggeized partons of the gauge--invariant Lipatov's Effective Field
Theory (EFT)~\cite{Lipatov:1,Lipatov:2},
which guarantees gauge--invariant definition of the hard scattering
coefficient ${\cal H}(x_i, t_i, \mu^2)$. Feynman rules for the EFT were
formulated in Ref.~\cite{Lipatov:3}, all needed for the present studies
rules are collected in Ref.~\cite{NS:2016} and implemented in the
{\tt ReggeQCD} model--file by M.~Nefedov for {\tt FeynArts}~\cite{FeynArts}.

We use model to obtain uPDFs from the collinear ones
firstly proposed by Kimber--Martin--Ryskin--Watt (KMRW)~\cite{KMR,MRW},
but with significant modifications explained in the Ref.~\cite{NS:2020}:
$$
\Phi_i(x, t, \mu^2) =
\frac{\alpha_S(t)}{2 \pi} \ \frac{T_i(x, t, \mu^2)}{t} \ \sum_j
\int_x^{\Delta(t, \mu^2)} dz \
P_{i j}(z) \ F_j\left( \frac{x}{z}, t \right),
$$
where $\Delta(t, \mu^2) = \mu / \left( \mu + \sqrt{t} \right)$ is
cutoff which ensures rapidity ordering and $F_i(x, \mu^2) = x \, f_i(x, \mu^2)$.
Function $T_i(x, t, \mu^2)$ usually called as a Sudakov form factor is based
on the KMRW model~\cite{KMR,MRW} and the condition of equivalence of the
exact normalization:
\begin{equation}
\int_0^{\mu^2} dt \ \Phi_i(x, t, \mu^2) = F_i(x, \mu^2) \label{eq2}
\end{equation}
and the definition:
$$
\Phi_i(x, t, \mu^2) = \frac{d}{dt} \left[ T_i(x, t, \mu^2) \ F_i(x, t) \right].
$$
The explicit form of the Sudakov form factor, which is depend on the $x$
oppositely to the original KMRW model, was first obtained in Ref.~\cite{NS:2020}.

The partonic subprocesses of the orders ${\cal O}(\alpha^1 \alpha_S^0)$ and
${\cal O}(\alpha^1 \alpha_S^1)$ which contribute to the direct photon
production are the following:
\begin{eqnarray}
Q \ (q_1) + \bar Q \ (q_2) & \to & \gamma \ (q_3), \label{pr1} \\
R \ (q_1) + Q \ (q_2) & \to & \gamma \ (q_3) + q \ (q_4), \label{pr2} \\
Q \ (q_1) + \bar Q \ (q_2) & \to & \gamma \ (q_3) + g \ (q_4), \label{pr3}
\end{eqnarray}
where $Q \ (\bar Q)$ and $R$ denote Reggeized quark (antiquark) and
Reggeized gluon respectively.

The LO partonic subprocess~(\ref{pr1}) corresponds to the case when there are
no other particles inside a photon cone
of the infinite radius $r^2 = \Delta y^2 + \Delta \phi^2 \to \infty$,
which is not exact suitable for the experimental setup, where $r$ is finite.
We will show below that this problem of the LO PRA can be solved.
The NLO${}^\star$ contributions from the subprocesses~(\ref{pr2}) and~(\ref{pr3})
should be added. The first one is infrared (IR) safe when integrating over
$|{\bf q}_{T_4}|$ down to zero since there is no the subprocess of a quark--gluon
scattering in the LO approximation. The subprocess~(\ref{pr3}) is IR diverge
and this divergence may be regularized only in the full NLO calculation including
loop corrections, see discussion in Ref.~\cite{NS:2017}.

\begin{figure}[ht]
\centering
\includegraphics[scale=0.7]{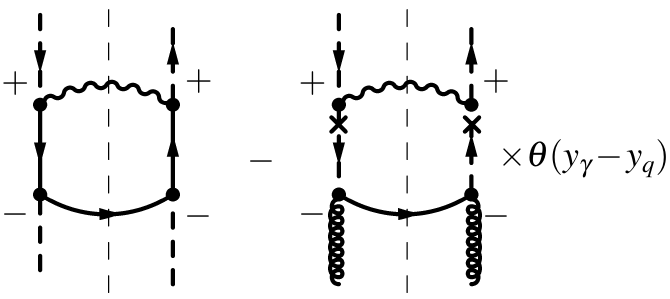}
\includegraphics[scale=0.7]{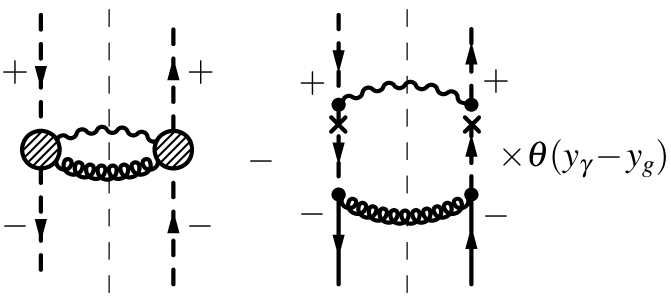}
\caption{The mMRK subtraction scheme for the
subprocesses~(\ref{pr2}) and~(\ref{pr3}).}
\label{fig1}
\end{figure}

Computations of the real emission NLO${}^\star$ corrections in the HEF have
non--trivial problems of double counting between tree--level corrections to
the hard scattering coefficient and uPDFs:
to calculate NLO${}^\star$ cross section, we must integrate over the entire
phase space of an additional parton $q$ in~(\ref{pr2}) and $g$ in~(\ref{pr3}),
including the $y_4 \in (- \infty, + \infty)$ integration. In the
PRA~\cite{NS:2020},
there are only three rapidity regions because of strong rapidty ordering:
forward (backward) rapidity region of the parton cascade with $y \to + (-) \infty$
of every parton, which consists partons from uPDF
$\Phi(x_{1 (2)}, t_{1 (2)}, \mu^2)$, and the
central region of the production particles with $- \infty < y < + \infty$.
But, when additional parton goes deeply to the forward (backward) rapidity
region, it should be included into the uPDF. To exclude such contribution and
then subtract, we put in the $t$--chanel amplitudes described
subprocesses~(\ref{pr2}) and~(\ref{pr3}) propagator to be Reggeized,
while $s$ and $u$ chanel amplitudes do not contribute to the subtraction term,
and consider the opposite to the photon (which is not connected by a vertex)
initial parton on mass--shell, so that the photon is produced in the central region.
A similar subtraction scheme was proposed earlier in Ref.~\cite{NS:2016}.
Subtraction terms for the subprocesses~(\ref{pr2}) and~(\ref{pr3}) are
demonstrated in Fig.~\ref{fig1}.

The subtraction amplitude for the subprocess~(\ref{pr2})
(with on--shell initial gluon)
from the Fig.~\ref{fig1} in the spinor representation has the form:
\begin{equation}
{\cal A}_a^{\mu \nu} = -i e g \, T_a \,
\bar u(q_4) \left( \gamma^\mu + \slashed{q} \frac{n_-^\mu}{q_2^-} \right)
D^+(q)
\left( \gamma^\nu - \slashed{q}_1 \frac{n_-^\nu}{q_3^-} +
\slashed{q} \frac{n_+^\nu}{q_3^+} \right) u(q_{L_1}), \label{eq:st}
\end{equation}
here $q = q_1 - q_3$,
$T_a$ is the ${\rm SU}(3)$ generators in the adjoint representation,
$q_{L_1}$ is the longitudinal part of the momenta $q_1$, on which the Reggeized
quark spinor depends.
The Reggeized quark propagator $D^\pm(q)$ applies the rapidity odering and has the
projection operator
$\hat P^\pm = \left( 1 / 4 \right) \slashed{n}^\mp \slashed{n}^\pm$
in the numerator~\cite{Lipatov:2}:
$$
D^\pm(q) =
\theta\left( y_3 - y_4 \right) \,
\frac{i\slashed{q}}{q^2} \, \hat P^\pm.
$$
It is easy to check
that amplitude~(\ref{eq:st}) satisfies Slavnov--Taylor identities for
real gluon and photon respectively due to gauge--invariance of the effective
vertices~\cite{Lipatov:1}:
$q_{2 \mu} {\cal A}_a^{\mu \nu} = 0$ and $q_{3 \nu} {\cal A}_a^{\mu \nu} = 0$.
After squaring and averaging over the spin and color indices, one can obtain:
$$
\overline{\mid {\cal A} \mid^2} =
32 \pi^2 \alpha \alpha_S \
\frac{C_F}{C_A (C_A^2 - 1)} \frac{x_1}{x_2}
\frac{\left( (q_3^-)^2 + (q_4^-)^2 \right) \left( q_3^- q_4^+ + \hat t - t_1 \right)}
{q_3^+ q_3^- \hat t},
$$
where $\hat t = (q_1 - q_3)^2$, $C_A = 3$ and $C_F = 4 / 3$.
Similarly, one can obtain subtraction term corresponding to the
matrix element for the subprocess~(\ref{pr3}).

To derive the formula for the subtraction term cross section,
one can integrate~(\ref{eq1}) over the $t_2$ with the help of~(\ref{eq2})
and take limit $t_2 \to 0$ in the hard scattering coefficient:
\begin{equation}
d \sigma_{\rm sub.} \simeq
\Phi(x_1, t_1, \mu^2) \otimes \lim\limits_{t_2 \to 0}{\cal H}(x_i, t_i, \mu^2)
\otimes f(x_2, \mu^2),
\label{eq3}
\end{equation}
such procedure is IR safety due to the correct collinear limit of the
modified MRK (mMRK) factorization used in the PRA~\cite{NS:2020}.

Numerical calculations of the cross sections are performed using the {\tt Suave}
Monte--Carlo algorithm implemented in the {\tt CUBA} library~\cite{Cuba}.

\section{Numerical results}
\label{sec:num}

\begin{figure}[t]
\centering
\includegraphics[scale=0.22]{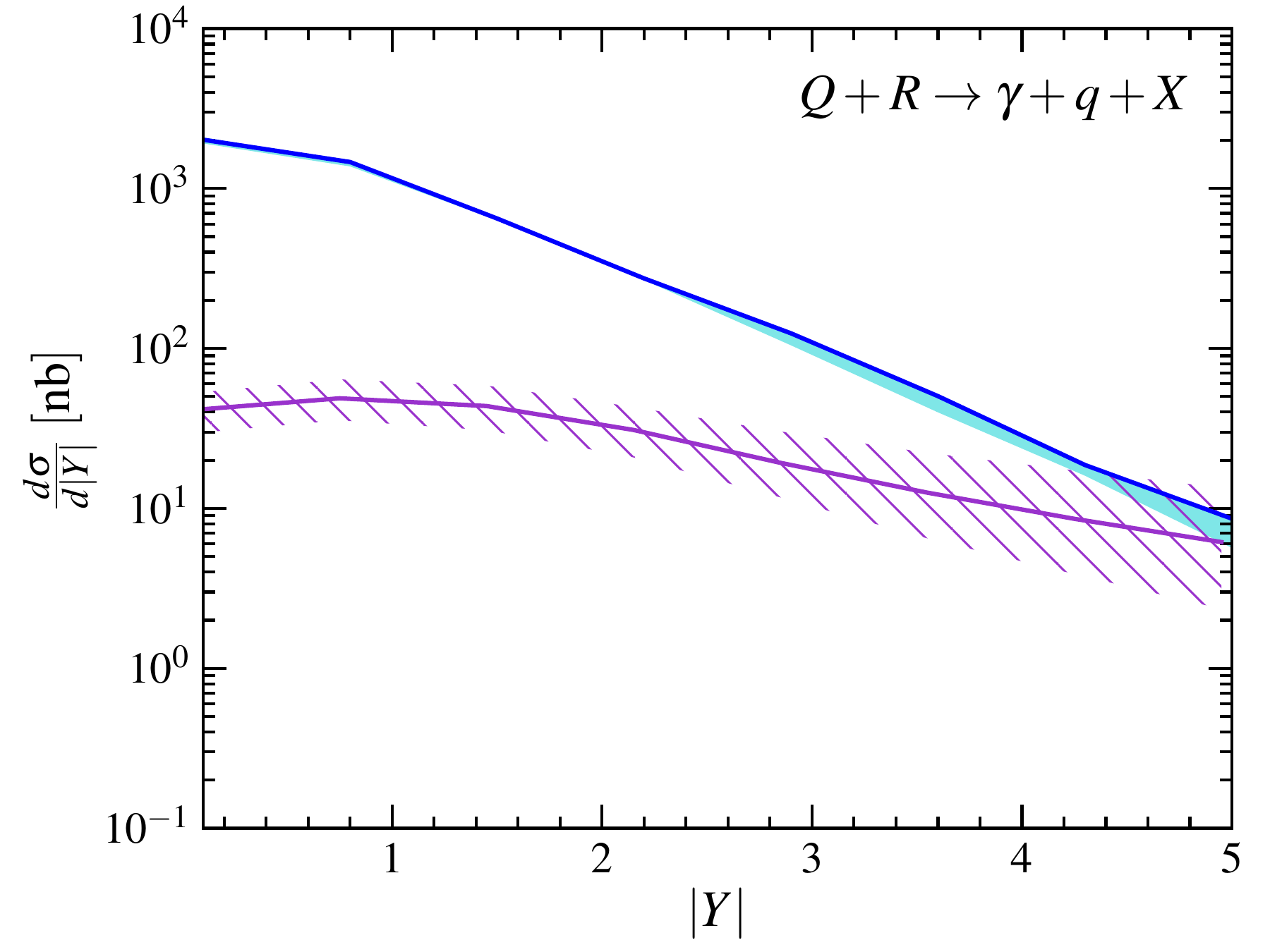}
\includegraphics[scale=0.22]{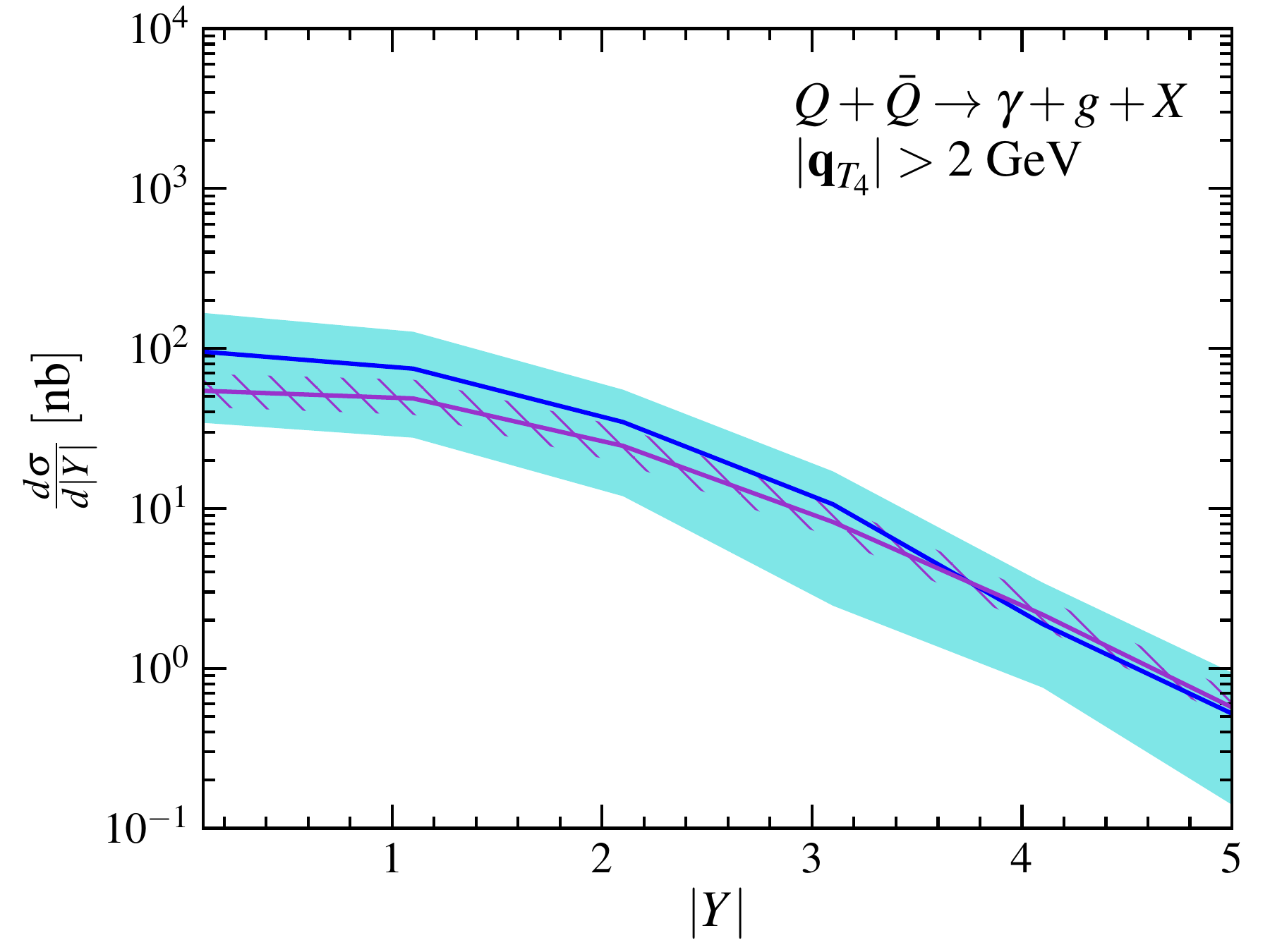}
\caption{Spectra for the rapidity difference $|Y| = |y_\gamma - y_{q, g}|$
of the subprocesses~(\ref{pr2}) and~(\ref{pr3}).
Solid lines are the unsubtracted contributions~(\ref{pr2}),~(\ref{pr3}),
and dashed are the relevant mMRK subtraction.}
\label{fig2}
\end{figure}

To suppress the contribution of fragmentation photons in the experiment,
an isolation condition is usually introduced:
\begin{equation}
r = \sqrt{\Delta y^2 + \Delta \phi^2} \geq r_0,\label{eq4}
\end{equation}
which means no hadrons with $E_T > E_T^{(\rm iso)}$ inside the photon isolation cone
$r \leq r_0$, where $r_0$ and $E_T^{(\rm iso)}$ are taken from the
experimental conditions~\cite{ATLAS:2016}.
To estimate the fragmentation photon contribution,
one can use Frixione modification of the isolation condition~\cite{Frixione:1998}:
\begin{equation}
E_T \leq E_T^{(\rm iso)} \ \chi(r ; n), \qquad
\chi(r ; n) =
\left( \frac{1 - \cos r}{1 - \cos r_0} \right)^n \label{eq5}
\end{equation}
with $n \geq 1 / 2$. In the measurements~\cite{ATLAS:2016},
$r_0 = 0.4$ and the $E_T^\gamma$--dependent isolation was used:
$$
E_T^{(\rm iso)} < 4.8 \ {\rm GeV} + 4.2 \cdot 10^{-3} \times E_T^{\gamma}.
$$
As we have found, the effect of the Frixione condition~(\ref{eq5}) is negligibly
small in the kinematical region of ATLAS measurements~\cite{ATLAS:2016}
and standard isolation cone condition is a good approximation of the
direct contribution we are interested in.

At first, we calculate spectra on the rapidity difference $Y = y_\gamma - y_{q, g}$
of the subprocesses~(\ref{pr2}) and~(\ref{pr3}) since they are the most
specific for the subtraction scheme, see Fig.~\ref{fig2}.
In the case of the subprocess~(\ref{pr3}), we numerically regularized the IR
divergence using cutoff $|{\bf q}_{T_4}| > 2$ GeV.
With growth of $|Y|$, mMRK subtraction contribution becomes large and coincides
with the unsubstracted cross section in the both cases. For the subprocess~(\ref{pr3}),
the subtraction term is large even at intermediate $|Y|$ and the resulting
contribution is small assuming independence from the cutoff. This fact shows
self--consistency of the PRA firstly demonstrated in Ref.~\cite{NS:2016}.

\begin{figure}[t]
\centering
\includegraphics[scale=.2]{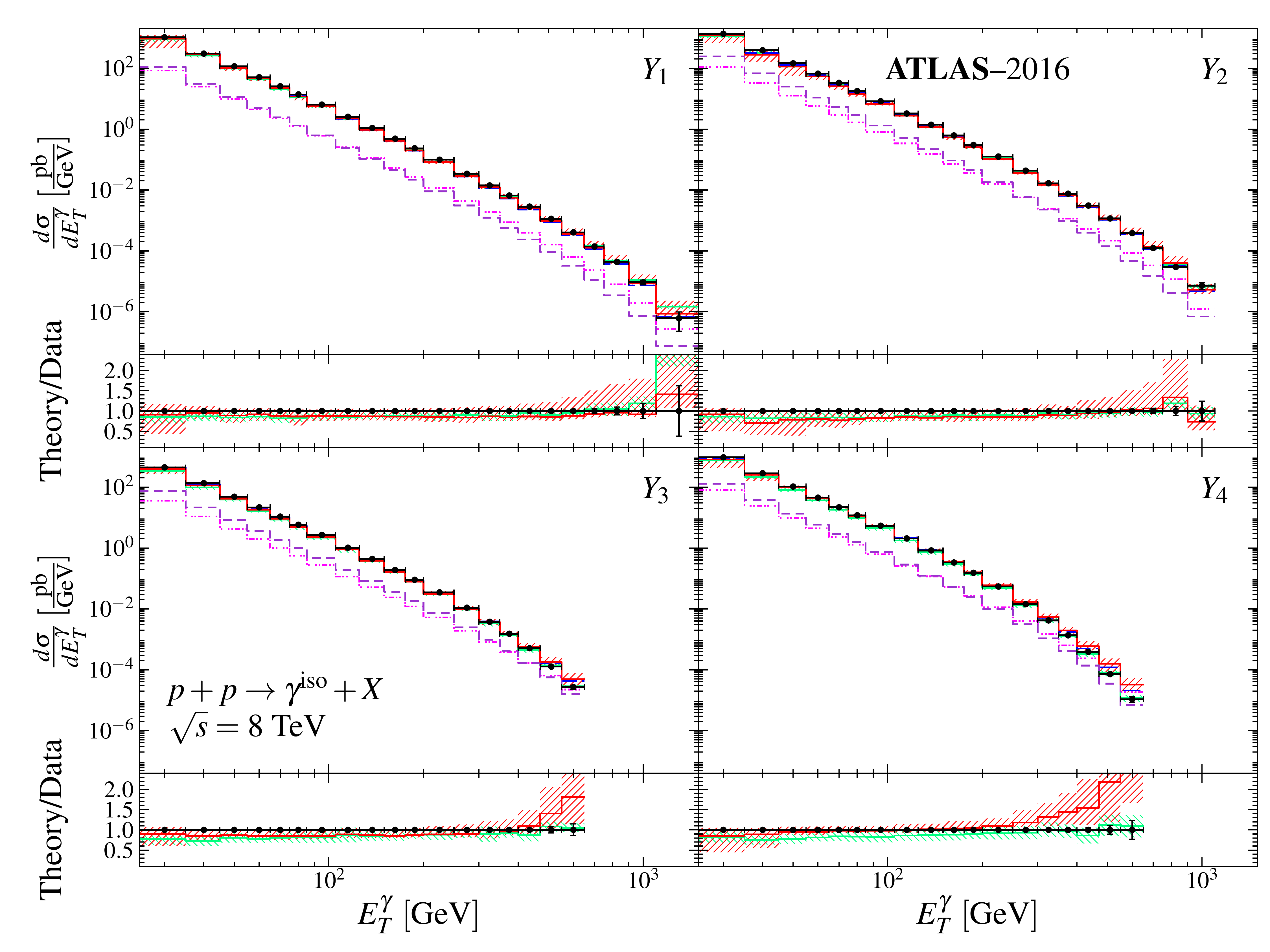}
\caption{Transverse energy spectra of the isolated photons at the $\sqrt{s} = 8$ TeV.
The LO~(\ref{pr1}) and NLO${}^\star$~(\ref{pr3}) contributions are shown separetly,
as well as mMRK subtraction term.
Solid red lines with hatches correspond to the total contribution after subtraction.
Solid green lines with corridor are the NLO CPM predictions obtained with
{\tt JetPhoX}~\cite{Catani:2002} parton level generator taken from
Ref.~\cite{ATLAS:2016}.}\label{fig3}
\end{figure}

We compare our predictions for the photon transverse energy
$E_T^\gamma$ spectra from $25$ GeV up to $1.1$ TeV with ATLAS data~\cite{ATLAS:2016}
at $\sqrt{s} = 8$ TeV in the four rapidity regions:
$Y_1$ -- $|y^\gamma| < 0.6$,
$Y_2$ -- $0.6 < |y^\gamma| < 1.37$,
$Y_3$ -- $1.56 < |y^\gamma| < 1.81$, and
$Y_4$ -- $1.82 < |y^\gamma| < 2.37$, see Fig.~\ref{fig3}.
Notations of the curves are explained in the figure caption.
As we can see in Fig.~\ref{fig3}, mMRK subtraction term mostly
coincides with the LO contribution~(\ref{pr1}) and only the NLO${}^\star$
term~(\ref{pr2}) remains. The agreement of our predictions with data is well,
exclude last large $E_T^\gamma$ bins in the $Y_4$ region where our
predictions overstimate experimental data by about a factor of $2$, because the
mMRK condition $\mu \ll \sqrt{s}$ becomes under the question in this region.

We also compare our results with the NLO CPM predictions as it was obtained
in Ref.~\cite{ATLAS:2016} with the help of parton level
event generator {\tt JetPhoX}~\cite{Catani:2002}.
The NLO CPM predictions are shown in Fig.~\ref{fig3}.
One can conclude that our results obtained using
NLO${}^\star$ approximation of the PRA agree well with NLO CPM calculation
as it should be for hard single--scale processes

\section{Conclusions}
\label{sec:con}
In the present paper, the single isolated photon production is studied in the
framework of the NLO${}^\star$ approximation of the PRA. The new mMRK double
counting between tree--level corrections and uPDFs subtraction scheme is proposed.
We have obtained a quite satisfactory description of $E_T^\gamma$ spectra
at LHC energies in the various rapidity regions. Additionally, we have shown that
mMRK subtraction term is large in case of NLO${}^\star$ subprocess
$Q \bar Q \to \gamma g$ where the LO subprocess $Q \bar Q \to \gamma$ exists,
which shows self--consistency of the PRA.

\section*{Acknowledgments}
We are grateful to M.~Nefedov for a fruitful discussion on the
subtraction scheme. The work is supported by the Foundation for the
Advancement of Theoretical Physics and Mathematics BASIS, grant No.
24--1--1--16--5 and by the grant of the Ministry of Science and High
Education of Russian Federation , No. FSSS--2024--0027.

\bibliographystyle{pepan}
\bibliography{pepan_biblio}

\end{document}